\documentclass[a4paper,12pt,reqno]{article}
\usepackage{amsmath}
\allowdisplaybreaks

\renewcommand{\i}{\mathrm{i}}
\renewcommand{\H}{\mathcal{H}}
\renewcommand{\L}{\mathcal{L}}
\newcommand{\al}{\alpha}
\newcommand{\ad}{\dot{\alpha}}
\newcommand{\be}{\beta}
\newcommand{\bd}{\dot{\beta}}

\newcommand{\de}{\delta}

\newcommand{\ep}{\varepsilon}
\newcommand{\si}{\sigma}
\newcommand{\bsi}{\bar{\sigma}}
\newcommand{\la}{\lambda}
\newcommand{\bla}{\bar{\lambda}}
\newcommand{\bpsi}{\bar{\psi}}
\newcommand{\half}{\tfrac{1}{2}}
\newcommand{\ihalf}{\tfrac{\mathrm{i}}{2}}
\newcommand{\quart}{\tfrac{1}{4}}

\newcommand{\tab}{\quad\,}
\newcommand{\bZ}{\bar{Z}}
\newcommand{\p}{\partial}
\newcommand{\T}{\mathcal{T}}
\newcommand{\D}{\mathcal{D}}
\newcommand{\Db}{\bar{\mathcal{D}}}
\newcommand{\sphi}{\varphi}
\newcommand{\bsphi}{\bar{\varphi}}
\newcommand{\bF}{\bar{F}}
\newcommand{\Chi}{\raisebox{2pt}{$\chi$}}
\newcommand{\bchi}{\raisebox{2pt}{$\bar{\chi}$}}
\newcommand{\com}[2]{[\,#1\, ,\,#2\,]}
\newcommand{\aco}[2]{\{#1\, ,\,#2\}}
\newcommand{\tbs}[1]{\overset{\leftrightarrow}{#1}}
\DeclareSymbolFont{AMSb}{U}{msb}{m}{n}
\DeclareMathSymbol{\fieldC}{\mathalpha}{AMSb}{"43}
\DeclareMathSymbol{\fieldR}{\mathalpha}{AMSb}{"52}
\DeclareMathSymbol{\fieldZ}{\mathalpha}{AMSb}{"5A}

\begin{document}


{\tiny Institut f\"ur Theoretische Physik, Universit\"at Hannover\hfill
       Institut f\"ur Theoretische Physik, Universit\"at Hannover} \medskip

\setlength{\unitlength}{.8mm}

\begin{center}
 \begin{picture}(21,16)
  \multiput(10,0)(7,0){2}{\framebox(4,16){}}
  \multiput(4,0)(1,0){2}{\line(0,1){16}}
  \multiput(0,16)(5,0){2}{\line(1,0){4}}
  \put(14,8){\line(1,0){3}}
  \put(4,16){\oval(8,32)[bl]}
  \put(5,16){\oval(8,32)[br]}
 \end{picture}
\end{center}

\vspace{-14mm}

ITP--UH--29/97 \\ hep-th/9711025 \hfill November 1997


\vspace{8mm}
\begin{center}
 {\Large\bfseries Gauging the Central Charge} \\[5mm]
 Norbert Dragon and Ulrich Theis \\[2mm]
 \textit{Institut f\"ur Theoretische Physik, Universit\"at Hannover \\
	Appelstra\ss{}e 2, 30167 Hannover, Germany}
\end{center}
\vspace{6mm}

\centerline{\bfseries Abstract} \medskip

We gauge the central charge of the $N=2$ supersymmetry algebra in rigid
superspace. The Fayet Sohnius hypermultiplet with gauged central charge
coincides on-shell with $N=2$ supersymmetric electrodynamics. The gauge
couplings of the vector-tensor multiplet turn out to be nonpolynomial.
\vspace{10mm}


Since the appearance of the vector-tensor multiplet \cite{SSW} in string
theory \cite{Louis} (see also \cite{Siebelink}), much effort has been made
to derive consistent interactions both with itself and with background
vector multiplets \cite{Claus1,Grimm,Kuzenko}. The presence of Chern-Simons
couplings in particular seemed to be a general feature of the possible
interactions.

Recently, however, a gauge theory was constructed outside the framework of
supersymmetry which resembled the vector-tensor multiplet with local
central charge transformations and resulted in a nonpolynomial action
without any Chern-Simons forms \cite{Brandt}. The same model was found by
different means as a particular example in a more general class of theories
(eq.~(17) of \cite{Henneaux}) where it is formulated in a polynomial
first-order form. In the present paper we obtain the supersymmetric
generalization of this action.

Gauging the central charge does not imply gravitational interactions. This
is remarkable because the transformation generated by the central charge is
frequently considered as a translation $\de_z$ in an additional dimension of
space-time and gauging translations leads to gravitational interactions.
Nevertheless the Fayet Sohnius hypermultiplet and the vector-tensor multiplet
with gauged central charge exhibit some features which are reminiscent of
gravity, in particular the inverse of some fields has to exist just as the
inverse of the metric in gravity.
\bigskip

Our considerations are based on the supersymmetry algebra
 \begin{align}\label{N2}
  \aco{\D_\al^i}{\Db_{\ad j}} & = -\i \de_j^i\, \si^a_{\al\ad} \D_a &
	\com{\D_a}{\D_b} & = F_{ab}\, \de_z \notag \\
  \aco{\D_\al^i}{\D_\be^j} & = \ep_{\al\be}\, \ep^{ij} \bZ \de_z &
	\com{\D_\al^i}{\D_a} & = \ihalf (\si_a \bla^i)_\al \de_z \\
  \aco{\Db_{\ad i}}{\Db_{\bd j}} & = \ep_{\ad\bd}\, \ep_{ij} Z \de_z &
	\com{\Db_{\ad i}}{\D_a} & = \ihalf (\la_i \si_a)_{\ad} \de_z\ .
	\notag
 \end{align}
The real generator $\de_z$ is central and commutes with all generators of
the supersymmetry algebra,
 \begin{equation}
  \de_z=\de_z{}^\ast\, ,\quad  \com{\de_z}{\D_\al^i} = 0\, ,\quad
	\com{\de_z}{\Db_{\ad}^j} = 0\, ,\quad \com{\de_z}{\D_a} = 0\, .
 \end{equation}
$\D_a$ is the gauge covariant derivative
 \begin{equation} \label{deriv}
  \D_a = \p_a + A_a \de_z
 \end{equation}
and $F_{ab}$ the abelian field strength of the gauged central charge
 \begin{equation}
  F_{ab} = \p_a A_b - \p_b A_a\ .
 \end{equation}
The algebra \eqref{N2} might as well be read as an algebra with a gauged
abelian transformation and no central charge. However, the scalar field $Z$
has to have a nonvanishing vacuum expectation value as we will see.

$Z$, $\bZ$, $\la_{\underline{\al}}^i$ and $A_a$ are component fields of the
vector multiplet which is completed by an auxiliary triplet $Y^{ij}=Y^{ji}
=(Y_{ij})^\ast$
 \begin{equation} \begin{split}
  \la_\al^i & = \D_\al^i Z\, ,\quad \bla_{\ad i} = \Db_{\ad i} \bZ\, , \\
  Y^{ij} & = \quart (\D^i \D^j Z + \Db^i \Db^j \bZ)\, , \\
  F_{ab} & = \quart (\D^i \si_{ab} \D_i Z - \Db_i \bsi_{ab} \Db^i \bZ)\, .
 \end{split} \end{equation}
They are tensor fields and carry no central charge. So they are unchanged by
gauged central charge transformations $s_z$ with a real gauge parameter
$C(x)$ with arbitrary space-time dependence. Under these transformations the
vector field $A_a$ is changed by the gradient of the gauge parameter,
 \begin{equation}
  s_z Z = 0\, ,\quad s_z \la_\al^i = 0\, ,\quad s_z \bla_{\ad i} = 0\, ,
  \quad s_z Y^{ij} = 0\, ,\quad s_z A_a = -\p_a C\, .
 \end{equation}
\bigskip

To construct invariant actions, we make use of the linear multiplet with
gauged central charge as in \cite{vanHolten}. One starts from a field
 \begin{equation}
  \L^{ij} = \L^{ji} = (\L_{ij})^\ast
 \end{equation}
satisfying the constraints
 \begin{equation} \label{linear}
  \D_\al^{(i} \L^{jk)} = 0 = \Db_{\ad}^{(i} \L^{jk)}.
 \end{equation}
It contains among its components a real vector $V^a = - \frac{\i}{6} \D_i
\si^a \Db_j \L^{ij}$, which is con\-strained by
 \begin{equation}
  \D_a V^a = \tfrac{1}{12} \de_z  \big[ Z \D_i \D_j + 3 Y_{ij} + 4 \la_i
	\D_j \big] \L^{ij} + \text{h.c.}
 \end{equation}
From the identity
 \begin{equation*}
  s_z (A_a V^a) = - (\p_a C) V^a + A_a C \de_z V^a = - \p_a (C V^a) +
	C \D_a V^a
 \end{equation*}
it follows immediately that
 \begin{equation} \label{L1}
  \L = \tfrac{1}{12} \big[ Z \D_i \D_j + 3 Y_{ij} + \i A_a \D_i \si^a
	\Db_j + 4 \la_i \D_j \big] \L^{ij} + \text{h.c.}
 \end{equation}
is invariant under gauged central charge transformations upon integration
over spacetime. This action is also N=2 supersymmetric, as can be checked
by explicit calculation. So once we have constructed a multiplet on which
the algebra \eqref{N2} is realized we try to construct a composite field
$\L^{ij}$ which satisfies \eqref{linear} and use \eqref{L1} as a Lagrangian.
\bigskip


\textbf{The Fayet Sohnius Hypermultiplet with Gauged Central Charge} \medskip

This multiplet is obtained from a complex doublet $\sphi^i$, $\bsphi_i =
(\sphi^i)^\ast$ subject to the constraints
 \begin{equation}
  \D_\al^{(i} \sphi^{j)} = 0 = \Db_{\ad}^{(i} \sphi^{j)}\, .
 \end{equation}

The independent components are
 \begin{equation} \label{comp}
  \sphi^i\, ,\quad \Chi_\al = \half \D_{\al i} \sphi^i\, ,\quad \bpsi_{\ad}
	= \half \Db_{\ad i} \sphi^i\, ,\quad F^i = \de_z \sphi^i
 \end{equation}
and their complex conjugates. The algebra \eqref{N2} determines their
supersymmetry transformations\footnote{We exchange without further notice a
vector index $a$ for a pair of spinor indices e.g.\ $\D_{\al\ad} =
\si_{\al\ad}^a \D_a$.}
 \begin{align}
  \D_\al^i \Chi_\be & = - \ep_{\al\be} \bZ F^i\, , & \quad
	\Db_{\ad}^i \Chi_\al & = \i \D_{\al\ad} \sphi^i\, , \notag \\
   \Db_{\ad}^i \bpsi_{\bd} & = \ep_{\ad\bd} Z F^i\, , & \quad \D_\al^i
	\bpsi_{\ad} & = -\i \D_{\al\ad} \sphi^i\, , \\
  \D_\al^i F^j & = \ep^{ij} \de_z \Chi_\al\, , & \quad \Db_{\ad}^i
	F^j & = \ep^{ij} \de_z \bpsi_{\ad}\, , \notag
 \end{align}
and their transformations under $\de_z$
 \begin{align}
  \de_z \Chi_\al & = - \frac{1}{Z} \big( \i \D_{\al\ad} \bpsi^{\ad} +
	\la_{\al i} F^i \big)\, , \notag \\[6pt]
  \de_z \bpsi_{\ad} & = - \frac{1}{\bZ} \big( \i \D_{\al\ad} \Chi^\al +
	\bla_{\ad i} F^i \big)\, , \label{deltaz} \\[6pt]
  \de_z F^i & = \frac{1}{|Z|^2} \big( \D^a \D_a \sphi^i + \la^i
	\de_z \Chi + \bla^i \de_z \bpsi - Y^{ij} F_j \big)\, . \notag
 \end{align}

The above central charge transformations $\de_z$ are manifestly covariant.
They are, however, defined only implicitly because the covariant
derivatives $\D_a$ on the right hand side also contain $\de_z$,
\eqref{deriv}. Solving for $\de_z$ of the fields as in \cite{Brandt} we find
 \begin{subequations}
 \begin{align}
  \de_z \Chi_\al & = -(|Z|^2 - A^a A_a)^{-1} \big[ \bZ (\i \p_{\al\ad}
	\bpsi^{\ad} + \la_{\al i} F^i) + \i A_{\al\ad} (\i \p^{\ad\be}
	\Chi_\be - \bla^{\ad}_i F^i) \big]\, , \label{de_zChi} \\[6pt]
  \de_z \bpsi_{\ad} & = -(|Z|^2 - A^a A_a)^{-1} \big[ Z (\i \p_{\al\ad}
	\Chi^\al + \bla_{\ad i} F^i) + \i A_{\al\ad} (\i \p^{\bd\al}
	\bpsi_{\bd} - \la^\al_i F^i) \big]\, , \\[6pt]
  \begin{split}
  \de_z F^i & = (|Z|^2 - A^a A_a)^{-1} \big[\, \p^a \p_a \sphi^i + \p_b A^b
	F^i + 2 A^b \p_b F^i - Y^{ij} F_j \\*
	    & \phantom{= (Z\bZ - A^a A_a)^{-1} \big[} + \la^i \de_z \Chi +
	\bla^i \de_z \bpsi\, \big]\, , \label{de_zF}
  \end{split}
 \end{align}
 \end{subequations}
which requires $(|Z|^2 - A^a A_a)$ not to vanish, i.e.\ $Z$ has to have a
nonvanishing vacuum expectation value. Explicit calculation confirms
that these expressions indeed transform as tensors under gauged central
charge transformations $s_z$, i.e.\ without derivatives of the transformation
parameter $C$.
\bigskip

Invariant actions are now easily found by constructing linear multiplets
out of the components \eqref{comp}. Since the constraints on $\sphi^i$ have
not changed as compared to ungauged central charge transformations, we can
use the composite fields already given by Sohnius \cite{Sohnius},
 \begin{equation}
  \L_0^{ij} = \sphi^{(i} \tbs{\de_z} \bsphi^{j)}\, ,\quad
	\L_1^{ij} = - 2\i \kappa\, \sphi^{(i} \bsphi^{j)}\, .
 \end{equation}
The first one leads to the following Lagrangian
 \begin{equation}
  \L_0 = \p^a \bsphi_i\, \p_a \sphi^i - \ihalf \big( \Chi \si^a \tbs{\p_a}
	\bchi + \psi \si^a \tbs{\p_a} \bpsi \big) + (|Z|^2 - A^a A_a)
	F^i \bF_i\, .
 \end{equation}
This result could have almost been anticipated in view of the central charge
transformation eq.~\eqref{de_zF}. Although the corresponding action is gauge
invariant we have no gauge interactions but a free massless theory because
the gauge field couples only to the auxiliary fields. The path integral,
however, is different from a free theory because it is restricted by
$(|Z|^2 - A^a A_a)>0$.

Interactions and masses are introduced by $\L_1^{ij}$ which yields the
Lagrangian
 \begin{equation} \begin{split}
  \frac{1}{\kappa} \L_1 & = \i A^a \big( \bsphi_i \tbs{\p_a}
	\sphi^i \big) + \i (|Z|^2 - A^a A_a)\, (\sphi^i \bF_i - \bsphi_i
	F^i) - \i Y^{ij} \bsphi_i \sphi_j \\*
  & \tab + \i \bsphi_i\, (\la^i \Chi + \bla^i \bpsi) - \i \sphi^i\, (\bla_i
	\bchi - \la_i \psi) + \i (Z \Chi \psi - \bZ \bchi \bpsi) \\*
  & \tab + A_a (\Chi \si^a \bchi - \psi \si^a \bpsi)\, .
 \end{split} \end{equation}
It is remarkable that here interactions follow from the linear multiplet
$\L_1^{ij}$, which in the case of global central charge transformations
gives only the mass terms while couplings to gauge fields of a possible
additional symmetry are tied to the kinetic energies through the covariant
derivatives.

From $\L = \L_0 + \L_1$ we now compute the Euler-Lagrange derivatives of
the hypermultiplet,
 \begin{align}
  \frac{\hat{\p}\L}{\hat{\p}\bF_i} & = (|Z|^2 - A^a A_a)\, (F^i + \i
	\kappa \sphi^i)\, , \notag \\
  \begin{split}
  \frac{\hat{\p}\L}{\hat{\p}\bsphi_i} & = - \p^a \p_a \sphi^i + 2\i\, \kappa
	A^a \p_a \sphi^i + \i \kappa\, \sphi^i \p_a A^a - \i \kappa\, Y^{ij}
	\sphi_j \\*
  & \tab - \i \kappa\, (|Z|^2 - A^a A_a) F^i + \i \kappa\, (\la^i \Chi + \i
	\bla^i \bpsi)\, ,
  \end{split} \\
  \frac{\hat{\p}\L}{\hat{\p}\bchi_{\ad}} & = - \i (\p^{\ad\al} - \i \kappa\,
	A^{\ad\al}) \Chi_\al - \i \kappa\, \bZ \bpsi^{\ad} - \i \kappa\,
	\sphi^i \bla^{\ad}_i\, , \notag \\
  \frac{\hat{\p}\L}{\hat{\p}\psi^\al} & = - \i (\p_{\al\ad} - \i \kappa\,
	A_{\al\ad}) \bpsi^{\ad} + \i \kappa\, Z \Chi_\al + \i \kappa\,
	\sphi^i \la_{\al i}\, .\notag
 \end{align}
These equations of motion imply $F^i = - \i \kappa\, \sphi^i$, and the
gauged central charge transformations reduce on-shell to gauged U(1)
transformations. We obtain on-shell N=2 supersymmetric electrodynamics.
For example, if one uses the equations of motion the transformation
\eqref{de_zChi} simplifies to
 \begin{equation}
  \de_z \Chi_\al = - \i \kappa\, \Chi_\al\, .
 \end{equation}
Nevertheless the action is different from $N=2$ supersymmetric
electrodynamics. A surprising feature is that the coupling of the gauged
central charge vanishes if the mass of the Fayet Sohnius hypermultiplet
vanishes.
\bigskip


\textbf{The Vector-Tensor Multiplet with Gauged Central Charge} \medskip

For the vector-tensor multiplet gauging the central charge is slightly
more involved. This multiplet is obtained from a real scalar field $L$,
which for ungauged central charge satisfies the constraints
 \begin{equation}
   L = L^\ast\, ,\quad D_\al^{(i} \bar{D}_{\ad}^{j)} L = 0\, ,\quad
	D^{(i} D^{j)} L = 0\, .
 \end{equation}
Simply replacing the flat derivatives $D_\al^{i}$ and $\bar{D}_{\ad}^{j}$
with gauge covariant derivatives $\D_\al^{i}$ and $\Db_{\ad}^{j}$ leads to
inconsistencies that show up at a rather late stage of the process of
evaluating the algebra on the multiplet components. This problem can be
avoided, however, by a suitable change of the above constraints. They must
preserve the field content as compared to the case of ungauged central
charge. A consistent set of constraints for the vector-tensor multiplet
with gauged central charge is given by
 \begin{equation} \begin{split} \label{constraints}
  \D_\al^{(i} \Db_{\ad}^{j)} L & = 0\, , \\
  \D^{(i} \D^{j)} L & = \frac{2}{\raisebox{-1pt}{$\bZ - Z$}} \big( \D^{(i}
	Z \D^{j)} L + \Db^{(i} \bZ \Db^{j)} L + \half L \D^{(i} \D^{j)} Z
	\big)\, .
 \end{split} \end{equation}

Let us briefly comment on how these constraints were obtained. We considered
the following Ansatz, with the coefficients being arbitrary functions of $Z$
and $\bZ$,
 \begin{equation} \begin{split}
  \D_\al^{(i} \Db_{\ad}^{j)} L & = f\, \D_\al^{(i} Z \Db_{\ad}^{j)} L
	- \bar{f}\, \Db_{\ad}^{(i} \bZ \D_\al^{j)} L\, , \\
  \D^{(i} \D^{j)} L & = F\, \Db^{(i} \bZ \Db^{j)} L + G\, \D^{(i} Z \D^{j)}
	L + H L\, \D^{(i} \D^{j)} Z\, .
 \end{split} \end{equation}
The constraints have to satisfy the necessary consistency conditions
 \begin{equation}
  \D_\al^{(i} \D^j \D^{k)} L = 0\, ,\quad \Db_{\ad}^{(i} \D^j \D^{k)} L
	= \D^{(i} \D^j \Db_{\ad}^{k)} L\, ,
 \end{equation}
which yield a system of differential equations for the coefficient functions.
The first condition requires
 \begin{equation}
  \bar{f} F = 0\, ,\quad H = \half G\, ,\quad \p_Z G = \half G^2\, ,\quad
	\p_Z F = F (\half G - f)\, ,
 \end{equation}
while the second leads to
 \begin{gather}
  f = H - \half F\, ,\quad \p_Z f = f (G - f)\, ,\quad \p_{\bZ} H = \bar{f} H
	+ \half F \bar{H}\, , \notag \\
  \p_{\bZ} F = F (\half \bar{G} + \bar{f})\, , \quad \p_{\bZ} G - \p_Z \bar{f}
	= f \bar{f} + \half F \bar{F}\, .
 \end{gather}

We chose to investigate the case $F \neq 0$ in detail and postpone a
discussion of the solutions to $F = 0$ as they are much more involved.
If $F \neq 0$, we have $f = 0$ and $G = 2H = F$, where $F$ has to
satisfy the differential equations
 \begin{equation}
  \p_Z F = \half F^2\, ,\quad \p_{\bZ} F = \half F \bar{F}\, .
 \end{equation}
The general solution is
 \begin{equation}
  F(Z,\bZ) = \frac{2 \mathrm{e}^{-\i\sphi}}{\raisebox{-1pt}{$\mathrm{e}^{\i
	\sphi} \bZ - \mathrm{e}^{-\i\sphi} Z + \i r$}}\, ,\quad r,\sphi \in
	\fieldR\, .
 \end{equation}
The parameters can be removed by a redefinition
 \begin{equation*}
  Z \rightarrow \mathrm{e}^{\i\sphi} (Z + \ihalf r)\, ,\quad \D_\al^i
	\rightarrow \mathrm{e}^{-\i\sphi/2} \D_\al^i\, ,
 \end{equation*}
which eventually leads to the constraints given in eq.~\eqref{constraints}.

The other solutions to the consistency conditions are currently under
investigation. In particular one has to determine all the transformations
and to check whether they indeed represent the algebra. Unfortunately, we do
not know of any other and shorter way to determine whether the constraints
are consistent.

The constraints \eqref{constraints} are linear in the components of the
 vector-tensor multiplet. This implies that we will not encounter any
self-interactions, in particular no coupling to a Chern-Simons form of the
vector component can arise.
\bigskip

We now investigate the consequences of the constraints \eqref{constraints}.
The independent components of the multiplet can be chosen as
 \begin{gather}
  L\, ,\quad \psi_\al^i = \i \D_\al^i L\, ,\quad \bpsi_{\ad i} = -\i
	\Db_{\ad i} L\, ,\quad  U = \de_z L\, , \notag \\
  G_{\al\be} = \half \com{\D_\al^i}{\D_{\be i}} L\, ,\quad \bar{G}_{\ad\bd}
	= \half \com{\Db_{i\ad}}{\Db_{\bd}^i} L\, , \\
  W_{\al\ad} = -\half \com{\D_\al^i}{\Db_{\ad i}} L\, . \notag
 \end{gather}
In addition some abbreviations will prove useful in the following,
 \begin{gather}
  I = \mathrm{Im}\, Z\, ,\quad R = \mathrm{Re}\, Z\, ,\notag \\
  \Lambda_a = \half (\la^i \si_a \bpsi_i + \psi^i \si_a \bla_i)\, ,\quad
	\Sigma_{ab} = \i (\la_i \si_{ab} \psi^i - \bpsi_i \bsi_{ab} \bla^i
	)\, .
 \end{gather}
The supersymmetry transformations then read
 \begin{align}
  \D_\al^i U & = - \i \de_z \psi_\al^i\, , \notag \\
  \D_\al^i W_a & = \big( \i \bZ\, \si_a \de_z \bpsi^i + \half U \si_a
	\bla^i - \D_a \psi^i \big)_\al\, , \notag \\
  \D_\al^i G_{ab} & = \big( 2I\, \si_{ab} \de_z \psi^i + U \si_{ab} \la^i
	+ \i \ep_{abcd} \si^c \D^d \bpsi^i \big)_\al\, , \\
  \D_\al^i \psi_\be^j & = \ihalf \ep^{ij} (\ep_{\al\be} \bZ U + \si^{ab}
	{}_{\al\be} G_{ab}) + \frac{\i}{2I} \ep_{\al\be} (\la^{(i}
	\psi^{j)} - \bla^{(i} \bpsi^{j)} + \i Y^{ij} L)\, ,\notag \\
  \Db_{\ad}^i \psi_\al^j & = \half \ep^{ij} \si^a_{\al\ad} (\D_a L + \i
	W_a)\, .\notag
 \end{align}

The central charge transformations contain covariant derivatives similar to
the case of the hypermultiplet and may be solved in complete analogy, but it
is more advantageous to use the manifestly covariant expressions
 \begin{align}
  \begin{split}
  \de_z \psi^i & = \frac{\i}{Z} \big( \si^a \D_a \bpsi^i - \la^i U \big)
	+ \frac{\i}{2ZI} \Big[ \ihalf \bZ \la^i U - Y^{ij} \psi_j + \i
	\si^a \bpsi^i \p_a \bZ \\*
  & \tab + L \si^a \p_a \bla^i + \half (D_a L - \i W_a) \si^a \bla^i -
	F_{ab} \si^{ab} \psi^i - \ihalf G_{ab} \si^{ab} \la^i \\*
  & \tab - \frac{\i}{2I} \la_j (\la^{(i} \psi^{j)} - \bla^{(i} \bpsi^{j)}
	+ \i Y^{ij} L) \Big]\, ,
  \end{split} \notag \\
  \de_z \big[ I & W_a - L \p_a R - \Lambda_a \big] = \D^b G_{ab}\, , \\
  \de_z \big[ I & \tilde{G}_{ab} + R G_{ab} + L F_{ab} + \Sigma_{ab} \big]
	= - \ep_{abcd} \D^c W^d\, .\notag
 \end{align}
Here and later on we use a tilde to denote the dual
\begin{equation}
\tilde{G}_{ab} = \half \ep_{abcd} G^{cd}.
\end{equation}

The vector $W_a$ and the antisymmetric tensor $G_{ab}$ are subject to the
constraints
 \begin{align}
  \D^a & \big[ I W_a - L \p_a R - \Lambda_a \big] = \half F_{ab}
	G^{ab}\, , \notag \\
  \D^a & \big[ I \tilde{G}_{ab} + R G_{ab} + L F_{ab} + \Sigma_{ab} \big]
	= \tilde{F}_{bc} W^c\, .\notag
 \end{align}
With the help of the central charge transformations we obtain from these
covariant expressions the equations
 \begin{align}
  \p^a & \big[ I W_a - A^b G_{ab} - L \p_a R - \Lambda_a \big] = 0\, ,
	\label{dW} \\
  \p^a & \big[ I \tilde{G}_{ab} + R G_{ab} + \ep_{abcd} A^c W^d + L
	F_{ab} + \Sigma_{ab} \big] = 0\, , \label{dG}
 \end{align}
which identify the brackets as duals of the field strength of a 2-form
$K_{ab}$ and a 1-form $T_a$, respectively,
 \begin{gather}
  I W_a = \half \ep_{abcd} (\p^b K^{cd} - A^b \tilde{G}^{cd}) + L \p_a R
	+ \Lambda_a\, , \label{IW} \\
  I \tilde{G}_{ab} + R G_{ab} = \ep_{abcd} (\p^c T^d - A^c W^d) - L F_{ab}
	- \Sigma_{ab}\, . \label{IGRG}
 \end{gather}
We cannot read off $W_a$ and $G_{ab}$, however, since the equations are
coupled as in \eqref{deltaz}. Solving for $W_a$ and $G_{ab}$ we obtain the
nonpolynomial expressions
 \begin{align}
  W_a & = \frac{1}{IE} \big( \H_a - |Z|^{-2} A_a A^b \H_b + |Z|^{-2}
	\T_{ab} A^b \big)\, , \\
  G_{ab} & = \frac{1}{|Z|^2} \Big[ \T_{ab} - \frac{R}{IE} \ep_{abcd} A^c
	\big( \H^d + |Z|^{-2} \T^{de} A_e \big) - \frac{2}{E} A_{[a} \big(
	\H_{b]} + |Z|^{-2} \T_{b]c} A^c \big) \Big]\, .
 \end{align}
Here we used the abbreviations
 \begin{align}
  E & = 1 - |Z|^{-2} A^a A_a\, , \notag \\
  T_{ab} & = \p_a T_b - \p_b T_a\, , \notag \\
  \H_a & = \half \ep_{abcd}\, \p^b K^{cd} + L \p_a R + \Lambda_a\, , \\
  \T_{ab} & = I (T_{ab} + L \tilde{F}_{ab} + \tilde{\Sigma}_{ab}) + R
	(\tilde{T}_{ab} - L F_{ab} - \Sigma_{ab})\, .\notag
 \end{align}

It remains to give the transformations of the fundamental fields $T_a$ and
$K_{ab}$. As the brackets in eqs.~\eqref{IW} and \eqref{IGRG} already
indicate, the central charge transformations read
 \begin{equation}
  \de_z T_a = - W_a\, ,\quad \de_z K_{ab} = - \tilde{G}_{ab}\, ,
 \end{equation}
while the supersymmetry generators act as
 \begin{align}
  \D_\al^i T_a & = \big( A_a \psi^i - \half L \si_a \bla^i - \i \bZ\,
	\si_a \bpsi^i \big)_\al\, , \\
  \D_\al^i K_{ab} & = - 2\i\, \big( I \si_{ab} \psi^i + \half L \si_{ab}
	\la^i + A_{[a} \si_{b]} \bpsi^i \big)_\al\, .
 \end{align}
Closure of the algebra on these two potentials, however, requires to add
total derivatives of conveniently chosen fields to the transformations as
is already known from the case of ungauged central charge.
\bigskip

Next, we construct a linear multiplet with lowest component $\L^{ij}$.
Given the constraints \eqref{constraints} on $L$ it is easy to find a field
with the required properties,
 \begin{equation} \begin{split}
  \L^{ij} & = \i \big( \Db^{(i} L \Db^{j)} L - \D^{(i} L \D^{j)} L - L
	\D^{(i} \D^{j)} L \big) \\
  & = \i (\psi^i \psi^j - \bpsi^i \bpsi^j) - \frac{\i}{I} L \big(
	\la^{(i} \psi^{j)} - \bla^{(i} \bpsi^{j)} + \i Y^{ij} L \big)\, .
 \end{split} \end{equation}
Applying the rule \eqref{L1} we eventually obtain the Lagrangian
 \begin{equation} \begin{split}
  \L & = \half I \big( \p^a\! L\, \p_a L - W^a W_a + (|Z|^2 - A^a A_a) U^2
	\big) - \half L^2 \p^a \p_a I \\*
  & \tab - \quart G^{ab} \big( I G_{ab} - R \tilde{G}_{ab} + 4 A_{[a} W_{b]}
	\big) - \frac{1}{4I} Y^{ij} Y_{ij} L^2 \\*
  & \tab + \text{fermion terms}\, ,
 \end{split} \end{equation}
where we used the constraints \eqref{dW}, \eqref{dG} to combine terms into
a total derivative which was dropped thereafter. To compare with the action
found in \cite{Brandt}, we insert the expressions for $W_a$ and $G_{ab}$,
 \begin{equation} \begin{split}
  \L_\mathrm{bos} & = \half I \p^a\! L\, \p_a L - \half L^2 \p^a \p_a I +
	\half I (|Z|^2 - A^a A_a) U^2 - \frac{1}{4I} Y^{ij} Y_{ij} L^2 \\*
  & \tab - \frac{1}{4|Z|^4} \T^{ab} (I \T_{ab} - R \tilde{\T}_{ab}) -
	\frac{1}{2IE} (\H_a + |Z|^{-2} \T_{ab} A^b)^2 \\*
  & \tab + \frac{1}{2I|Z|^2 E} (A_a \H^a)^2\, .
 \end{split} \end{equation}
With $L$ and $U$ set to zero and $Z$ an imaginary constant, which breaks
supersymmetry but keeps central charge invariance, the bosonic Lagrangian
indeed coincides with the one given in \cite{Brandt}.

It would be interesting to include self-interactions as in \cite{Kuzenko}
by a further deformation of the above constraints. This may lead to a
superspace description of the results presented in \cite{Claus1}.
In addition there might be different ways to gauge the central charge
corresponding to different consistent constraints.
\bigskip

While this paper was in preparation similar results on multiplets with gauged
central charge and $N=2$ supergravity were published in \cite{Claus2}.
\bigskip

\textbf{Acknowledgement} \\
We thank S.~M.~Kuzenko for useful discussions.

\end{document}